\documentclass[preprint]{aastex}

\def\beqa{\begin{eqnarray}}
\def\eeqa{\end{eqnarray}}
\def\beq{\begin{equation}}
\def\eeq{\end{equation}}

\slugcomment{Submitted to Astrophysical Journal}

\shortauthors{Puchalla et al.}
\shorttitle{MAT-SOURCES}

\begin{document}

\title{Galactic Observations at 31, 42 and 144~GHz with the Mobile
Anisotropy Telescope} 

\author{ J.~L.~Puchalla\altaffilmark{1},
R.~Caldwell\altaffilmark{2},
K.~L.~Cruz\altaffilmark{1},
M.~J.~Devlin\altaffilmark{1},
W.~B.~Dorwart\altaffilmark{2},
T.~Herbig\altaffilmark{2},
A.~D.~Miller\altaffilmark{2},
M.~R.~Nolta\altaffilmark{2},
L.~A.~Page\altaffilmark{2},
E.~Torbet\altaffilmark{2},
H.~T.~Tran\altaffilmark{2} }

\altaffiltext{1}{University of Pennsylvania, Department of Physics and
Astronomy, David Rittenhouse Laboratory, Philadelphia, PA 19104 \hspace{.5in} 
e-mail contact: puchalla@higgs.hep.upenn.edu} 
\altaffiltext{2}{Princeton University, Physics Department, Jadwin Hall,
Princeton, NJ 08544}

\begin{abstract}

We present multi-frequency maps of a portion of the Galactic Plane
centered on a declination of $-60^{\circ}$ with resolutions ranging
from $0^{\circ}.2$ to $0^{\circ}.9$. The maps are optimized to detect
unresolved source emission and are cross-calibrated against the planet
Jupiter. We discuss six millimeter-bright regions, three of which are
visible in all bands, and list measured flux densities. Variability is
limited to less than 3.8\% for all sources seen at 31 and 42~GHz and less
than 10\% at 144~GHz. Fractional polarization limits smaller than 0.08 are
measured for all sources seen at 31 and 42~GHz. No fractional
polarization limits are reported at 144~GHz.

\end{abstract}


\section{Introduction}

Galactic radiation at centimeter and millimeter wavelengths is
attributed to a combination of the following sources of emission:
synchrotron, free-free, thermal interstellar dust
and variations in the electric and magnetic
dipoles of dust ({\it i.e.}  spinning dust). Large-area,
multi-frequency surveys help us to understand the objects and
environments which emit in the far-infrared and assist in determining
the mechanisms driving microwave emission. The atmosphere is
relatively opaque at these frequencies making such surveys difficult
to perform from low altitudes.  At present, the only comprehensive
microwave survey data available is from the Cosmic Background Explorer
({\it COBE}) \cite{ben96}. It will not be until the launch of the
Microwave Anisotropy Probe ({\it MAP}) satellite in late 2000 that a
full sky survey with pixel resolution smaller than $1^{\circ}$ will be
completed.

Bright millimeter-wavelength sources are important for cosmology as
calibrators and a possible Cosmic Microwave Background (CMB)
foreground contaminate \cite{toff99,sok98}. There are several open
questions about this class of object that include: what are their
spectra; are they variable; are they polarized?  We address these
questions for six Galactic sources using observations made with the
Mobile Anisotropy Telescope ({\it MAT}). We focus on characterizing
emission from unresolved and localized millimeter-bright
regions. These data will also help assess daytime atmospheric
conditions relevant to observations from similar Chilean sites.  We
present maps of the Galactic Plane centered on a declination of
$-60^{\circ}$ (approximately $280^{\circ}<l<305^{\circ}$ and
$-8^{\circ}<b<4^{\circ}$) covering several bright regions including
the Carinae Nebula and RCW~57.

All of the millimeter-bright objects we observe have {\it
IRAS} counterparts or are known {\it HII} regions.  The {\it IRAS}
survey indicates that most of these
contain substructure unresolved by the {\it MAT} observations. 
Since they cannot be uniquely labeled with previous
surveys we denote each bright region with number from 1 to
6. Figure~\ref{fig:region} shows the region covered by these observations. 
Table~\ref{table:sources}  lists some of the known radio and
millimeter-bright sources observed in other surveys which
comprise these regions. The list of associated names gives the
brightest far-infrared and radio sources in the vicinity.

\begin{table*}
\caption{Location of each of the {\it MAT} regions along with
known radio and infrared-bright objects contained therein. The approximate
location of the centroid is listed to a precision appropriate for all
{\it MAT} frequencies. MAT~5 and MAT~6 are sub-categorized for the 144~GHz 
channels since multiple sources are resolved. IRAS-PSC denotes the
{\it IRAS} point source catalog.}\label{table:sources}
\bigskip
\begin{centering}
\begin{tabular}{lcccc} \hline \hline
\multicolumn{1}{l}{MAT Object} & 
\multicolumn{1}{c}{Center RA} & 
\multicolumn{1}{c}{Center Dec} & 
\multicolumn{1}{c}{Associated Names} & 
\multicolumn{1}{c}{Type} \\ 
\hline
MAT 1 & $151^{\circ}.8$  &   $-57^{\circ}.3$   & IRAS 10049-5657 & IRAS-PSC \\ 
      &        &           & RCW46          & HII region \\

MAT 2 & $156^{\circ}.0$  &   $-57^{\circ}.8$  & RCW49 & HII Region \\ 
      &        &          & NGC 3247 & Open Cluster \\ 
MAT 3 & $161^{\circ}.0$  & $-59^{\circ}.7$  & $\eta$Carinae & Nebula \\ 
      &        &           & IRAS 10414-5919 &  \\
MAT 4 & $165^{\circ}.1$  & $-61^{\circ}.1$  & IRAS 10591-6040 & IRAS-PSC \\ 
      &        &           & IRAS 10589-6034 &  \\
MAT 5 & $168^{\circ}.6$  & $-61^{\circ}.3$     & RCW57          & HII Region\\ 
MAT 5a & $167^{\circ}.90$  & $-61^{\circ}.35$  & NGC 3576 & HII Region \\ 
MAT 5b & $168^{\circ}.7$  & $-61^{\circ}.30$   & NGC 3603 & OB Cluster \\ 
MAT 5b &       &          &          & HII Region \\ 
MAT 6 & $183^{\circ}.3$  & $-62^{\circ}.9$  &  & IRAS-PSC \\ 
MAT 6a & $182^{\circ}.48$  &   $-62^{\circ}.82$           & IRAS 12073-6233   &  \\
MAT 6b &  $183^{\circ}.84$  &   $-63^{\circ}.00$     & IRAS 12127-6244   &  \\
\hline 
\end{tabular}
\end{centering}
\end{table*}

\section{Instrument and Observations}

{\it MAT} is a 0.8~m diameter aperture telescope based on the design
in Wollack {\it et al.} (1997). Three frequency bands (31, 42, 144~GHz) in
eight channels are monitored.  The telescope is steerable in
azimuth but fixed in elevation. The focal plane contains five
corrugated feeds. Observations are made in two polarizations in
Ka-band (26-36~GHz) and Q-band (36-46~GHz) using total power high
electron mobility transistors (HEMT) amplifiers. Observations are also
made in D-band with two single polarization SIS receivers with dual
sidebands ($\approx$~138-140~GHz and $\approx$~148-150~GHz with the
local oscillator center at 144~GHz). The calibration and antenna
patterns are measured in the field using multiple observations of
Jupiter ($\theta_{disk} \approx$ 30''-45''). Table~\ref{table:beams}
lists the {\it MAT} beam characteristics for each frequency. Antenna
patterns are modeled as two-dimensional Gaussian profiles
parameterized by an amplitude and two orthogonal
widths\footnote{FWHM is related to $\sigma$ by
$\sigma=\Theta_{FWHM}/[2(2ln2)^{\frac{1}{2}}]$.} $\sigma_{H}$ and
$\sigma_{V}$. The H and V denote the horizontal (chop direction)
and vertical (meridian) line with respect to the local horizon.  Except for
one of the SIS channels (D2) not located in the center of the focal
plane, all antenna patterns are well approximated by Gaussian
profiles.

\begin{table*}
\caption{{\it MAT} beam characteristics. The D2
beam response is not well modeled as a Gaussian below
-10dB. }\label{table:beams}
\bigskip
\begin{centering}
\begin{tabular}{ccccc} \tableline \tableline
\multicolumn{1}{c}{Center Freq.} & 
\multicolumn{1}{c}{Channel} & 
\multicolumn{1}{c}{$\sigma_{H}$} & 
\multicolumn{1}{c}{$\sigma_{V}$} & 
\multicolumn{1}{c}{Solid Angle} \\ % [-.15in]
\multicolumn{1}{c}{(GHz)} & 
\multicolumn{1}{c}{} & 
\multicolumn{1}{c}{(deg)} & 
\multicolumn{1}{c}{(deg)} & 
\multicolumn{1}{c}{($\Omega \times10^{-5}$str)} \\ \tableline

31 &Ka1\&2  & 0.39  & 0.39 & 30  \\ 
42 &Q1\&Q2  & 0.28  & 0.28 & 15  \\ 
42  &Q3\&Q4  & 0.31  & 0.29 & 18 \\ 
144 &D1  & 0.086  & 0.082 & 1.4  \\ 
144  &D2  & 0.13  & 0.12  & 2.9  \\ 
\hline 
\end{tabular}
\end{centering}
\end{table*}

The center of the focal plane is fixed at an elevation of
$40^{\circ}.76$. A flat mirror sweeps the beams with an amplitude of
$3^{\circ}$ on the sky in a sinusoidal pattern at 3.7~Hz. When the
mirror is set to zero amplitude, the center of the focal plane is at an
azimuth of $207^{\circ}.41$ East of North. Each day, {\it MAT} views 
a band about the South Celestial Pole of
constant mean declination. To view calibrators and other
bright objects, the telescope base is slewed to a new
azimuth and the source drifts through the field of
view. This procedure typically requires 20 minutes.

Observations were carried out at 5200~m from Cerro Toco\footnote{The
Cerro Toco site of the Universidad Cat\'{o}lica de Chile was made
available through the generosity of Prof. Hern\'{a}n Quintana , Dept. of
Astronomy and Astrophysics. It is near the proposed MMA site.} at latitude
22\fdg95 S, longitude 67\fdg78 W.  For this paper we
present data obtained between August 30, 1998 and October 14,
1998. The primary goal of the experiment was to use nightly
observations far from the Galactic Plane ($b>15^{\circ}$) to measure
the anisotropy in the CMB radiation 
\cite{mil99,torb99,dod99}.  This observation scheme also
provides a daily view of a large section of the Galactic Plane.

\subsection{Calibration}

Jupiter's emission spectrum is complicated by several molecular lines.
However, for wideband observations such as those made by {\it MAT}, Jupiter is
stable and its intrinsic brightness temperature has been measured to
$\approx 5\%$ \cite{uli81,gri86,gold97}. We use the values 152~K,
160~K, and 170~K for the 31, 42 and 144~GHz bands respectively. 

Four terms contribute most to the uncertainty of the calibration of
each channel: (1) the standard deviation in the measured solid angle;
(2) the standard deviation in the measured temperature of Jupiter for
each channel; (3) the intrinsic uncertainty of the brightness
temperature of Jupiter; (4) the uncertainty due to calibration drift
between Jupiter observations.  For the SIS channel 
used in this
analysis, the first three terms contribute uncertainties of 5.5\%, 7\%, and 5\%, respectively. For
the HEMT channels, the corresponding values are less than 7\%, 10\%, and 5\%.
These terms add in quadrature. The fourth term is primarily the result of
diurnal temperature variations.

To monitor calibration changes between observations of Jupiter, the
HEMT-based channels are injected with a signal from a
thermally-stabilized noise source for 40~ms twice every 100
seconds. The SIS-based channels are coupled to a 149~GHz tone at an
effective temperature of $\sim 1~K$.  These data indicate a 5\% random
uncertainty in both the SIS and HEMT calibrations and a $<2.5\%$
gain change at the time of the Galactic observations. After
accounting for the gain change, we estimate the total calibration
uncertainty for each channel by adding the uncorrelated errors in
quadrature giving a 11\% error for the SIS channel and a 14\%
error for the HEMT channels. The uncertainty quoted in Miller {\it et
al.} is smaller than that stated here because 
there was little temperature change, and consequently little gain
drift, between the Jupiter observations and the CMB observations. 

\subsection{Pointing}

Pointing is established through both rising and setting observations
of Jupiter. A nominal central azimuth and beam elevation are assumed
initially.  A Global Positioning System (GPS)
time-stamp is stored with each data point and is used to fix the time and
location of the centroid of the measured antenna pattern for each
horn.  Ephemeris data are then used to calculate a position error.  
The nominal central azimuth and beam
elevation\footnote{Measurements of Galactic sources indicated a tilt
in the base of the telescope of $0^{\circ}.10$.}  are then adjusted to
minimize the variance between the expected and predicted location of
Jupiter.

This model allows an accurate reconstruction of pointing over the full
azimuthal range of the telescope for the data considered
here. After coadding the data for this analysis, the
effective center of each of the 5 horns' antenna patterns is
determined with an absolute accuracy of $0^{\circ}.045$ and a relative
accuracy of $<0^{\circ}.01$ degrees.

\section{Data Reduction}

We consider a subset of daily observations of the Galactic Plane
that contains several millimeter-bright
sources. Each of the 8 amplifier channels is processed independently.
We do not include the 144~GHz channel D2 since its larger beam greatly
decreases its sensitivity to unresolved sources.

Of the total collected Galactic data,
54\% are rejected due to signal saturation in one or more channels and
$<1\%$ are rejected  based on telemetry drop-outs and noise-injector
operation; this
procedure leaves approximately 15 observations of 2.68~hours each.

Data are coadded for each chopper cycle based on the position of the
chopping mirror. The chopper position over a cycle never deviates from
the average cycle for the entire observation region by more that
1\%. The 144~GHz channels are binned into samples of $0^{\circ}.045$
in azimuth for 40 chopper cycles (about 11 seconds). The Ka and Q
channels are binned into samples three times larger. By fixing the
number of chopper cycles coadded and knowing both the time associated
with each data point and the location of each horn antenna pattern
(from previous observations of Jupiter), we construct an unfiltered
map of the sky.

The quality of daytime observations at this time of year is degraded
from those at night. This effect is well correlated to the rise in local
temperature above $0^{\circ}$C.  We remove the contribution from
atmospheric emission to these Galactic observations using a purely
spatial filter.  After performing a $10\sigma$ cut on isolated points
(i.e. inconsistent with the point spread function) we remove a slowly
varying polynomial from each coadded azimuthal strip. In this technique, a
sliding box-car average is generated. Points greater than $2.5\sigma$
as well as the four nearest neighbors are flagged.  The sliding
box-car average is then regenerated with the flagged points removed
from consideration. The results are insensitive to small changes in
this flagging level. In all channels, the box-car is 2 times the size
of the FWHM of the antenna pattern for that channel. This second
average is subtracted from the unfiltered data.

Monte Carlo analysis on simulated maps constructed from measured
atmospheric data at this site show that this filter effects
measurements of the amplitude and width of a Gaussian beam by $<1\%$
for the atmospheric conditions encountered in this data set. The
largest adverse effects occur at the edge of the maps where incomplete
coverage of a source can be confused for an atmospheric
variation. This occurs with two sources for one feed horn (Q3\&4)
and, hence, the measured source amplitudes for these channels have
been omitted from all averages.

\section{Maps}

After filtering, each daily map is coadded with observations from
other days to produce final images of the region. Each map is
comprised of between 10 and 15 coadded viewings. No attempt has been
made to coadd maps made from different amplifier channels. We show
representative plots of the Galactic region at 31, 42 and 144~GHz in
Figure~\ref{fig:ka_region}, \ref{fig:q_region}, and \ref{fig:d_region}
respectively. Also shown are close-ups of each source region for the
frequencies at which they were observed.  

Each source close-up is comprised of color contours and an intensity
color-bar associated with the {\it MAT} data. {\it IRAS}\footnote{Released
data product at http://www.ipac.caltech.edu/} intensity contours
have also been superimposed to demonstrate relative pointing and additional
bright sources in the region not seen at all of the {\it MAT} frequencies.

We are able to place constraints on both polarization and variability of
those sources observed at 31 and 42~GHz. Though all detector channels
are polarized, the antenna response of orthogonal polarizations in Ka
and Q bands are matched. This allows the fractional polarization 
$2(E_{||}-E_{\perp})/(E_{||}+E_{\perp})$ over
the source to be constrained as well as test for systematic errors.
Table~\ref{table:polar} gives the measured fractional polarization
limits for each source over a $3\sigma$ region. All sources are unpolarized 
to the level of our sensitivity.

\begin{table*}
\caption{Fractional polarization limits at 31 and 42~GHz. Due to
location, MAT~1 and
MAT~2 are only observed by the 42~GHz channels.}
\label{table:polar}
\bigskip
\begin{centering}
\begin{tabular}{ccc} \hline \hline
\multicolumn{1}{c}{MAT} & 
\multicolumn{1}{c}{Freq.} & 
\multicolumn{1}{c}{Frac. Polarization} \\ % [-.15in]
\multicolumn{1}{c}{Object} & 
\multicolumn{1}{c}{(GHz)} & 
\multicolumn{1}{c}{Limit} \\ \hline
MAT 1 & 42  & $< 0.06 $\\
MAT 2 & 42  & $< 0.05 $\\
MAT 3 & 31 & $< 0.02 $\\
      & 42 & $< 0.01 $\\
MAT 4 & 31 & $< 0.03 $\\
      & 42 & $< 0.03 $\\
MAT 5 & 31 & $< 0.03 $\\
      & 42 & $< 0.04 $\\
MAT 6 & 31 & $< 0.02 $\\
      & 42 & $< 0.08 $\\
\hline 
\end{tabular}
\end{centering}
\end{table*}

Nearly all compact radio sources demonstrate some variability on times
scales ranging from a few days to years.  The variability in emission
for the {\it MAT} sources is constrained over the observing period in all
frequency bands. We place limits on the daily and
total variation of peak brightness temperatures for all sources
observed at 31 and 42~GHz except MAT~4. Due to the lower
signal-to-noise of MAT 4 and all sources at 144~GHz we group these
observations into 3 subsets in time; the limits are only on the
total variation. Daily variation is defined as the standard deviation
of all observations about a line. Total variation is the standard
deviation on the mean value of all observations.
Table~\ref{table:vary} shows the limits to which all sources are
stable over the observing period. No statistically significant change
in brightness is indicated over any time frame.

\begin{table*}
\caption{Limits on (D) daily and (T) total brightness temperature
variation. N.A. indicates that placing this limit is not possible
from these data.}
\label{table:vary}
\bigskip
\begin{centering}
\begin{tabular}{cccc} \hline \hline
\multicolumn{1}{c}{MAT} & 
\multicolumn{1}{c}{Freq.} & 
\multicolumn{2}{c}{$1\sigma$ Limit on Variability} \\ % [-.15in]
\multicolumn{1}{c}{Object} & 
\multicolumn{1}{c}{(GHz)} & 
\multicolumn{1}{c}{({\bf D}aily)} &
\multicolumn{1}{c}{({\bf T}otal)} \\ \hline
MAT 1 & 42  & D$<$13.4\% & T$<$3.8\% \\
MAT 2 & 42  & D$<$5.4\% & T$<$1.5\% \\
MAT 3 & 31 & D$<$1.6\% & T$<$0.5\% \\
      & 42 & D$<$3.0\% & T$<$0.8\% \\
      & 144 & N.A. & T$<$8.5\% \\
MAT 4 & 31 & N.A. & T$<$7.5\%  \\
      & 42 & D$<$5.5\% & T$<$1.5\% \\
MAT 5 & 31 & D$<$4.3\% & T$<$1.2\% \\
      & 42 & D$<$7.2\% & T$<$2.0\% \\
a     &144 & N.A. & T$<$8.6\% \\
b     &144 & N.A. & T$<$21.4\% \\
MAT 6 & 31 & D$<$4.6\% & T$<$1.3\% \\
      & 42 & D$<$8.5\% & T$<$2.4\% \\
a     &144 & N.A. & T$<$10\% \\
b     &144 & N.A. & T$<$10\% \\
\hline 
\end{tabular}
\end{centering}
\end{table*}

\section{{\it MAT} Sources: Discussion}

The six {\it MAT} sources are extended at $100\mu$m in the {\it IRAS}
maps and cannot be unambiguously resolved in any of the {\it MAT}
beams.  For this reason, the flux density from each identified bright
region is determined within a circle whose radius is the average
$1\sigma$ and $2\sigma$ beam width for that frequency. That is, the
flux density at frequency $\nu$ is:
\beq
S_{\nu}=\int_{n\sigma} \frac{2kT(x,y)\nu^{2}}{c^{2} 10^{-26}} dx dy
\hspace{.25in} {\rm Jy}
\eeq
where $n\sigma$ denotes the area (number of $\sigma$) over which to
integrate and T(x,y) is the measured temperature profile over this
region. We also quote the amplitude ({\it i.e.} peak brightness
temperature) of the best-fit Gaussian profile (widths unconstrained)
for each source.  Table~\ref{table:howbright} details these results.

We average the measurements from all channels of a given
frequency. The results at 42~GHz are rescaled to a single beam size of
$\Omega_{42GHz}=16.5\times10^{-5}$str before averaging.  Due to the
low signal-to-noise, Q2 is not included in the 42~GHz average.

The quoted uncertainty results from the quadrature sum of the
uncorrelated components of the error divided by the
number of channels coadded. In addition to calibration error, the sum
includes an uncertainty in the baseline of the filtered maps of 4\%
for the SIS channel and 3\% for the HEMT channels.

\begin{table*}
\caption{Peak brightness temperature and flux density $S_{\nu}$
within a circle of $1\sigma$ and $2\sigma$ beam widths for the {\it
MAT} sources. No entry for a channel implies the source is not seen or
is below the measurable flux density limit of the map. We report the
average measured values for each frequency independent of
polarization. Quoted errors do not include the 5\% intrinsic
uncertainty in the brightness temperature of Jupiter. The map limit is
the flux density
at which an unresolved source would be measured with a signal-to-noise
of one.}\label{table:howbright}
\bigskip
\begin{centering}
\begin{tabular}{lccccc} 
\hline \hline
\multicolumn{1}{c}{MAT} & 
\multicolumn{1}{c}{Freq. Band} & 
\multicolumn{1}{c}{Peak Temp} & 
\multicolumn{1}{c}{ $1\sigma$-$S_{\nu}$} &
\multicolumn{1}{c}{ $2\sigma$-$S_{\nu}$} &
\multicolumn{1}{c}{Map Limit (Jy)} \\ % [-.15in]
\multicolumn{1}{c}{Object} & 
\multicolumn{1}{c}{(GHz)} & 
\multicolumn{1}{c}{(mK)} & 
\multicolumn{1}{c}{(Jy)} &
\multicolumn{1}{c}{(Jy)} &
\multicolumn{1}{c}{(Jy)} \\ 

\hline

MAT 1& &   &  & \\ 
&42  & 7.2 $\pm$ 0.7  & 22.2 $\pm$ 2.1  &   60.0$\pm$5.7 & 5.2 \\ \hline
MAT 2 & &  &  &\\ 
&42  & 26.6 $\pm$ 2.5  & 85.0$\pm$8.1   &    240.1$\pm$22.8 & 5.2 \\  \hline 

MAT 3 & &  &  & \\
&31  & 65.0 $\pm$ 6.2 & 220.1$\pm$20.9   &   539.2$\pm$51.2 &  7.1 \\  
&42   & 46.3 $\pm$ 3.6  & 178.1$\pm$13.9   &   496.9$\pm$38.8 &  5.2 \\  
&144   & 8.8 $\pm$ .9  &  33.1$\pm$3.7   &    110.6$\pm$11.9 & 12.8 \\  \hline

MAT 4 & &  &  & \\  
&31  & 12.0 $\pm$ 1.1  & 34.1$\pm$3.2  &  83.1$\pm$7.9 & 7.1 \\  
&42   & 7.6 $\pm$ 0.7  & 25.5$\pm$2.4   &  67.6$\pm$6.4 & 5.2 \\  \hline

MAT 5 & &  &  & \\  
&31  & 40.7 $\pm$ 3.9 &  140.8$\pm$13.4  &  341.6$\pm$32.5 & 7.1 \\  
&42   & 28.7 $\pm$ 2.2  &  101.5$\pm$7.9  & 244.9$\pm$19.1 & 5.2 \\  
a&144  & 8.8 $\pm$ 0.9  & 30.6$\pm$3.4    &   67.1$\pm$7.3 & 12.7\\ 
b&144  & 12.4 $\pm$ 1.37  & 46.4$\pm$5.1    &   117.8$\pm$12.7 & 12.7\\  \hline

MAT 6 & &  &  & \\ 
&31  & 17.4 $\pm$ 1.7  &  62.3$\pm$5.9  &  165.5$\pm$15.7 & 7.1 \\  
&42   & 8.8 $\pm$ 1.7  &  24.8$\pm$4.6   &    49.4$\pm$9.4 &  5.2 \\  
a&144  & 3.1 $\pm$ 0.7  & 10.9$\pm$2.6  &   34.2$\pm$8.5  & 12.7 \\ 
b&144  & 2.5 $\pm$ 0.7  &  10.2$\pm$2.6 &  28.9$\pm$7.2 & 12.7 \\  \hline

\hline 
\end{tabular}
\end{centering}
\end{table*}

MAT~1 is an extended HII region containing a single $100\mu$m-bright
{\it IRAS} point source. MAT~2 is a compact HII region. Since they are
observed only at 42~GHz we cannot constrain the spectral indices.
MAT~4 is also an extended HII region containing two moderately bright
$100\mu$m-bright {\it IRAS} point sources. It is seen at 31 and 42~GHz
but is not measurable at 144~GHz. The implied spectral index based on
the low frequency observations is $\alpha = -0.72$ where $S \sim
\nu^{\alpha}$. In this frequency range, the emission is dominated by
synchrotron radiation.

MAT~3 is dominated by emission from the Keyhole Nebula. This region
contains the known millimeter-bright and variable remnant
$\eta$Carinae.  Cox {\it et al.} (1995) have stated that at
wavelengths greater than 1~cm, $\eta$Carinae is dominated by free-free
emission but that between 1~cm and 1~mm the spectral index rises to
$\sim 1$. Cox {\it et al.} also give evidence of time variability in a
one year period of $30\%$.  Since our beam size at 144~GHz is
significantly larger, we are not able to constrain the emission of
$\eta$Carinae but rather the emission from the larger Keyhole Nebula
in which it is contained. This region also contains a second source
(IRAS 10414-5919) known to be bright at both 4.8~GHz \cite{con95} and
$100\mu$m. Therefore, there is no reason to expect that our map would
be centered on $\eta$Carinae.  Given the extended emission of this
region we have not attempted to constrain the spectral index.

MAT~5 and MAT~6 both contain two {\it IRAS}-bright sources which are clearly
seen at 144~GHz. However, the 31 and 42~GHz observations indicate that
only one source dominates emission at lower frequencies implying a highly
inverted spectrum. For MAT~6a, if we use the flux limit at 42~GHz in
conjunction with the measured flux at 144~GHz assuming it is an
unresolved source we find a spectral index of $\alpha \ge 1.5$ where
$S \sim \nu^{\alpha}$. For MAT~5a, we find $\alpha \ge 2.0$. Both are
consistent with the classical spectra of compact HII regions where $S
\sim \nu^{2}$.

\acknowledgments

We would like to thank Suzanne Staggs, Dave Rusin, Simon Dicker, Jeff
Klein, and Neill Reid for helpful conversations and assistance in
preparing this manuscript.  This work was supported by an NSF NYI
award, a Cottrell Award from the Research Corporation, a David and
Lucile Packard Fellowship (to LP), a NASA GSRP fellowship to AM, an
NSF graduate fellowship to MN, a NSF Career award (AST-9732960, to
MD), NSF grants PHY-9222952, PHY-9600015, AST-9732960, and the
University of Pennsylvania.  The data will be made public upon
publication of this article.

\figcaption[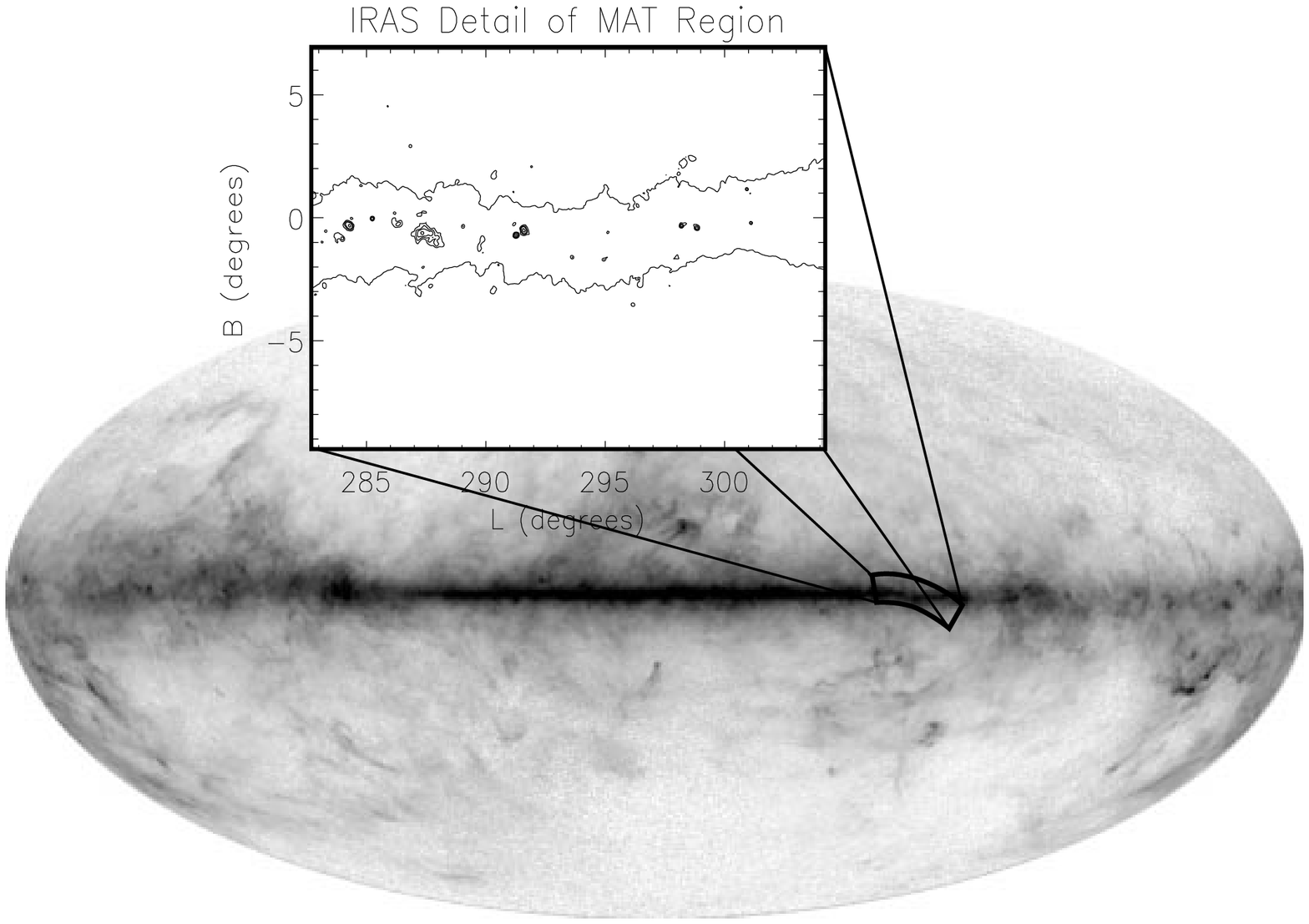]
{The location of the {\it MAT} observations
reported herein superimposed on the DIRBE $240\mu$m full sky map
\cite{leis96}. The inset details this region showing $100\mu$m bright
contours as seen by {\it IRAS}. \label{fig:region}}

\figcaption[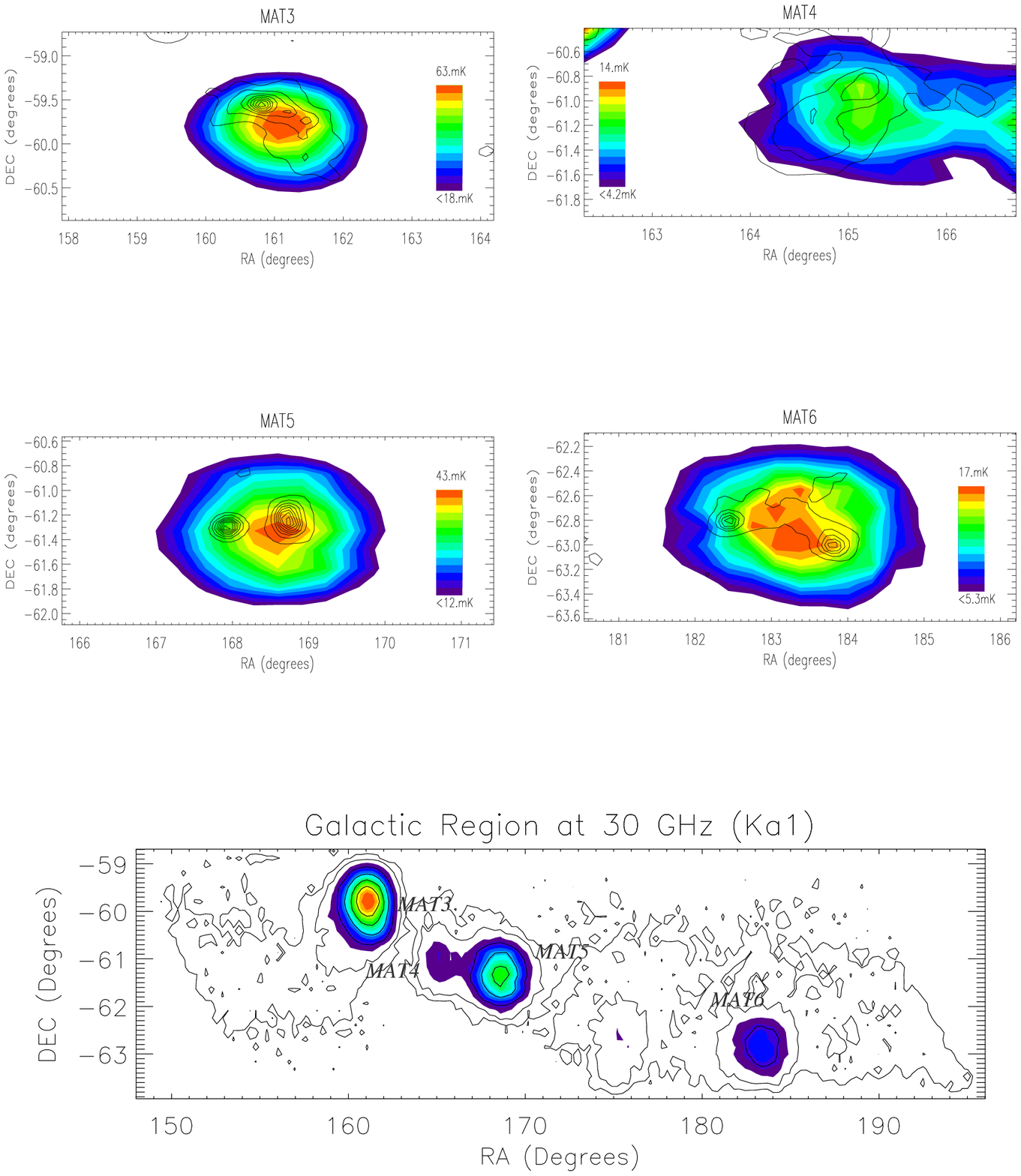]
{Observed sources at 31~GHz with {\it IRAS} contours and full 
Galactic region map measured by Ka1. The contours in the lower map are from the
{\it MAT} data. \label{fig:ka_region}}

\figcaption[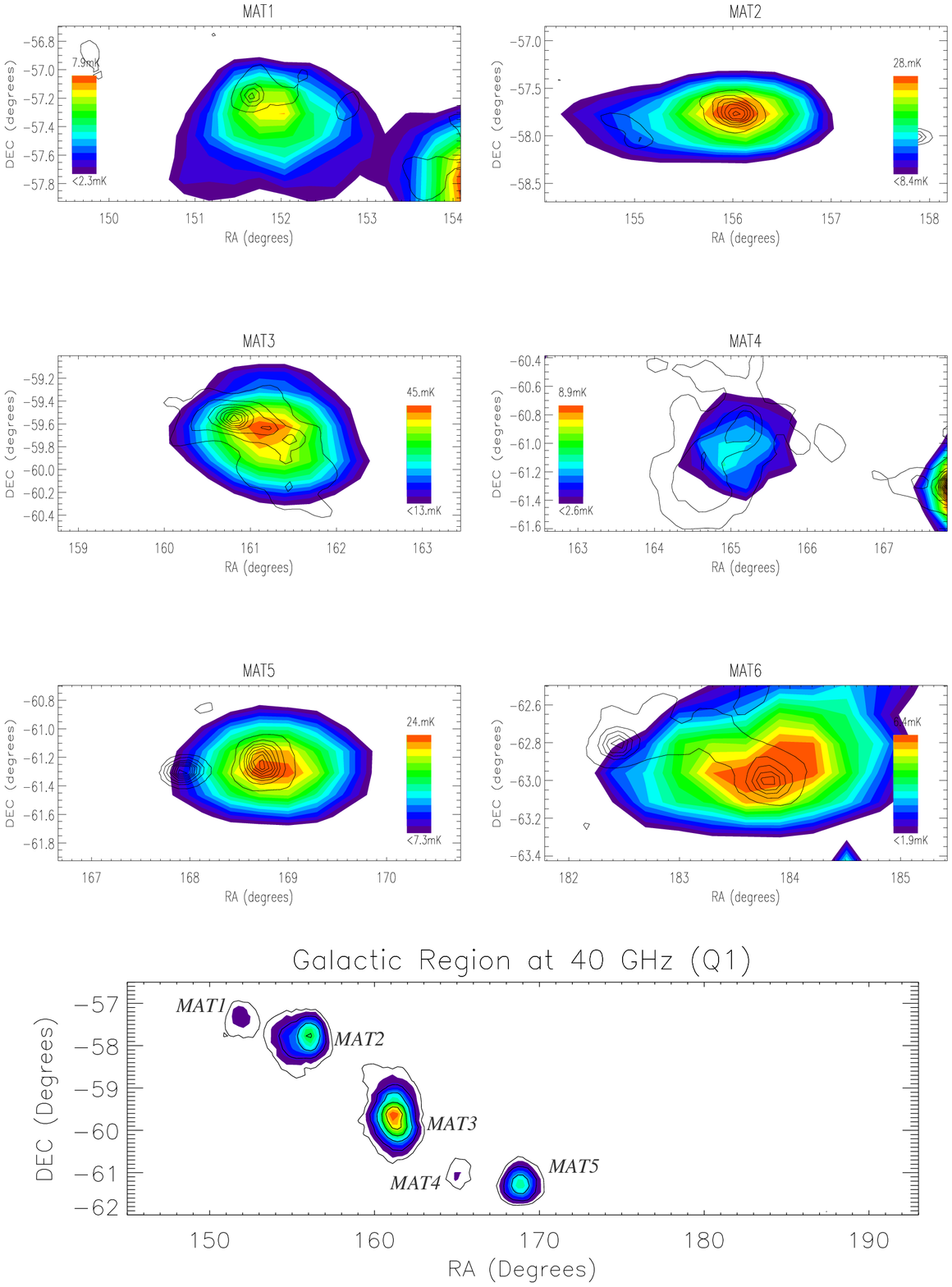]
{Observed sources at 42~GHz with {\it IRAS} contours and full 
Galactic region map measured by Q1.
MAT6 is not observed by this channel. The contours in the lower map are from the
{\it MAT} data. \label{fig:q_region}}

\figcaption[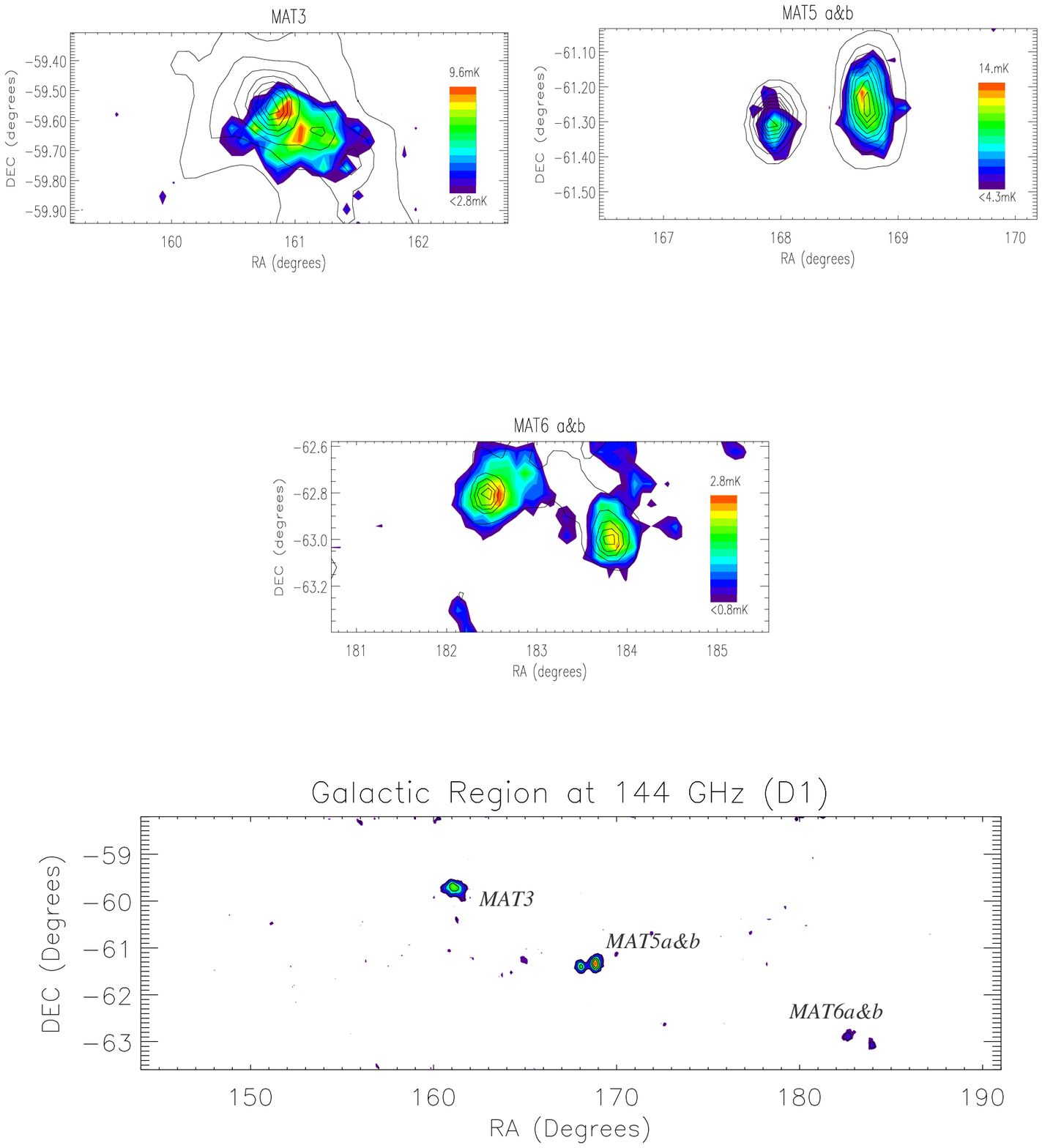]
{ Observed sources at 144~GHz with {\it IRAS} contours and
full Galactic region map measured by D1. MAT1 and MAT2 is out of the
field of view of D1. MAT4 is below the D1 sensitivity limit. The contours
in the lower map are from the {\it MAT} data. \label{fig:d_region}}

\end{document}